\documentclass[final,letter]{siamltex}

\usepackage{graphicx}
\usepackage{amsmath}
\usepackage{amssymb}
\usepackage{subfigure}

\providecommand{\eq}{\triangleq}
\DeclareMathOperator*{\argmin}{arg\,min}

\newcommand{\field}[1]{\mathbb{#1}}
\newcommand{\R}{\field{R}}
\newcommand{\N}{\field{N}}

\newcommand{\E}{{\mathcal{E}}}
\newcommand{\T}{\top}
\newcommand{\vc}[1]{{\boldsymbol{#1}}}
\newcommand{\Bs}{B}

\newcommand{\vx}{\vc{x}}
\newcommand{\vu}{\vc{u}}
\newcommand{\vz}{\vc{0}}

\newtheorem{thm}{Theorem}[section]

\title{Sparsely-Packetized Predictive Control by\\Orthogonal Matching Pursuit
	\thanks{This research was supported in part under 
	the MEXT (Japan) Grant-in-Aid for Young Scientists (B) No.~22760317,
	and also Australian Research Council's Discovery Projects funding scheme
	(project number DP0988601).}
}

\author{Masaaki Nagahara%
		\thanks{Graduate School of Informatics, Kyoto
    		University, Kyoto, 606-8501, 
    		Japan; e-mail: \texttt{nagahara@ieee.org}} \and
    Daniel~E.~Quevedo%
    		\thanks{School of Electrical Engineering \& Computer Science, 
		The University of Newcastle, NSW 2308, Australia; 
		e-mail:\texttt{dquevedo@ieee.org}} \and
	Jan~\O stergaard%
		\thanks{Department of Electronic Systems, 
		Aalborg University, Denmark; 
		e-mail: \texttt{janoe@ieee.org}}
}

\begin{document}

\maketitle



\pagestyle{empty}
\thispagestyle{plain}
\markboth{M.~Nagahara, D.~E.~Quevedo, and J.~\O stergaard}{Sparsely-Packetized Predictive Control by OMP}

\begin{abstract}
We study packetized predictive control,
known to be robust against packet dropouts in
networked systems. To obtain sparse packets for rate-limited networks,
we design control packets via an $\ell^0$ optimization, which can be effectively solved by 
orthogonal matching pursuit. Our formulation ensures asymptotic stability of
the control loop  in the presence
of bounded packet dropouts. 
\end{abstract}

\section{Packetized Predictive Control}
\thispagestyle{empty}

\label{sec:plant-model-control}

\begin{figure}[t]
\centering%
\includegraphics[width=0.6\linewidth]{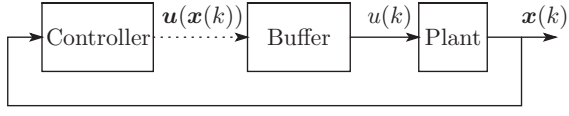}
\caption{Networked Control System with PPC}
\label{fig:NCS}
\end{figure}

Let us consider the following 
discrete-time, LTI plant model with a scalar input:
\begin{equation}
  \label{eq:plant}
  \vx(k+1)=A\vx(k)+\Bs u(k), \quad k\in\N_0, \quad \vx(0)=\vx_0,
\end{equation}
where $\vx(k)\in\R^n$ and $u(k)\in\R$ for $k\in\N_0$.
We assume that $(A,\Bs)$ is reachable.
We are interested in a networked control architecture where the controller
communicates with the plant actuator through an erasure channel, see Fig.~\ref{fig:NCS}.
This channel introduces bounded
packet-dropouts.
In packetized predictive  control (PPC), as
described, e.g., in \cite{QueNes11,NagQue11}
 at each time instant $k$, the controller uses the state $\vx(k)$ of the plant
\eqref{eq:plant} to calculate and
send a control packet of the form
\begin{equation}
  \label{eq:packet}
  \vu\bigl(\vx(k)\bigr)\eq
    \bigl[
	u_0\bigl(\vx(k)\bigr), u_1\bigl(\vx(k)\bigr),\dots, u_{N-1}\bigl(\vx(k)\bigr)
    \bigr]^\T\in\R^N
\end{equation}
to the plant input node. 

To achieve robustness
against packet dropouts, buffering is used.
To be more specific, suppose that at time instant $k$
the data packet $\vu(\vx(k))$ defined in \eqref{eq:packet} is successfully received at
the plant input side. Then, this packet is stored in a buffer, overwriting its
previous contents.
If the next packet, $\vu(\vx(k+1))$, is dropped,
then the plant input $u(k+1)$ is set to $u_1(\vx(k))$, the second element of
$\vu(\vx(k))$.
The subsequent elements of $\vu(\vx(k))$ are then successively used until
some control packet $\vu(\vx(k+\ell))$, $\ell\geq 2$,
is successfully received, i.e., no dropout occurs.

\section{Sparse Control Packet Design for Asymptotic Stability}
\label{sec:design}

In the PPC method discussed above,  control packets $\vu(\vx(k))$ are transmitted
at each time $k\in\N_0$ through an erasure channel.
It is often the case that the bandwidth of the channel is limited,
and hence one has to compress control packets to a smaller data size.
To achieve an efficient compression of packets,
we adopt a technique developed in
{\em compressed sensing} \cite{Don06}
that considers the {\em sparsity} of a signal (or a vector).
A sparse vector contains many $0$-valued elements and
can be highly compressed by only coding its few nonzero components.
Based on this notion, in \cite{NagQue11} we presented a sparsity-inducing
$\ell^1/\ell^2$ optimization for PPC which gives  \emph{practical} stability, i.e., existence of a
bounded invariant set. In the present work, we 
embellish the approach of  \cite{NagQue11} by showing how to find 
{\em sparse} control packets $\vu(\vx(k))$ ensuring that
the networked system with bounded packet dropouts is
{\em asymptotically stable}.
For this, we propose to design the control packets via the following
sparsity-promoting optimization:
\begin{equation}
 \vu(\vx(k))\eq \argmin_{\vu\in\R^N} \|\vu\|_0
 ~\text{subject to}~
  \|\vx_{N|k}\|_P^2 + \sum_{i=0}^{N-1}\|\vx_{i|k}\|_Q^2 \leq \vx(k)^\top W \vx(k),
 \label{eq:opt0}
\end{equation}
where $\|\vu\|_0$ is the number of the nonzero elements in 
$\vu = [u_0,u_1,\ldots,u_{N-1}]^\top$ and
\begin{equation}
   \vx_{0|k} = \vx,~\vx_{i+1|k} = A\vx_{i|k} + \Bs u_i,~i=0,1,\ldots,N-1.
   \label{eq:const2}
\end{equation}
In~(\ref{eq:const2}), $N$ is the horizon length, taken here equal to the buffer
size. The matrix $Q>0$ is a design parameter which allows one to shape the
response of 
the plant state components; $P>0$ and $W>0$ are chosen such that the feedback system is
asymptotically stable, as indicated in Theorem~\ref{thm:stability} below. To
state our subsequent results, we introduce the matrices $\Phi$, $\Upsilon$ and $\bar{Q}$ by
\[
  \Phi \eq \begin{bmatrix}\Bs & 0 & \ldots & 0\\
  			A\Bs & \Bs & \ldots & 0\\
			\vdots & \vdots & \ddots & \vdots\\
			A^{N-1}\Bs & A^{N-2}\Bs & \ldots & \Bs
			\end{bmatrix},~
  \Upsilon \eq \begin{bmatrix}A\\A^2\\\vdots\\A^N\end{bmatrix},~
  \bar{Q} \eq \mathrm{blockdiag}\{\underbrace{Q,\ldots,Q}_{N-1}, P\}.
\]
\begin{thm}[Asymptotic Stability]
\label{thm:stability}
Assume that the number of consecutive packet-dropouts
is uniformly bounded by $N$.
Suppose the matrices $P$, $Q$, and $W$
are chosen by the following procedure:
\renewcommand{\theenumi}{\roman{enumi}}
\begin{enumerate}
\item Choose $Q>0$ arbitrarily.
\item Solve the following Riccati equation to obtain $P>0$:
\begin{equation}
 P = A^\top P A - A^\top P\Bs(\Bs^\top P\Bs)^{-1}\Bs^\top PA + Q.
 \label{eq:ric}
\end{equation}
\item Compute the constants $\rho\in[0,1)$ and $c>0$ via
\[
 \rho \eq 1-\lambda_{\min}(QP^{-1}),~ 
 c \eq \frac{1-\rho^N}{1-\rho}\max_{i} \lambda_{\max} \bigl\{\Phi_i^\top P\Phi_i(\Phi^\top\bar{Q}\Phi)^{-1}\bigr\},
\]
where $\Phi_i$ is the $i$-th column of matrix $\Phi$.
\item Choose a matrix $\E$ such that $0<\E<(1-\rho)P/c$.
\item Set $W:=P-Q + \E$.
\end{enumerate}
Then for any choice of the control vector 
from the feasible set of the optimization
\eqref{eq:opt0} (which includes 
the sparsest control packets $\vu(\vx(k))$),
the networked control system is asymptotically stable,
that is,
$\lim_{k\rightarrow\infty}\vx(k)=\vz$.
\end{thm}

\section{Orthogonal Matching Pursuit}
\label{sec:OMP}

The optimization \eqref{eq:opt0} can be rewritten as follows:
\begin{equation}
  \vu\bigl(\vx(k)\bigr)\eq\argmin_{\vu\in\R^N} \|\vu\|_0~~ \text{subject to}~~ \|G\vu-H\vx(k)\|_2^2 \leq \vx(k)^\top W \vx(k),
  \label{eq:opt0_2}
\end{equation}
where $G \eq \bar{Q}^{1/2}\Phi$ and  $H \eq -\bar{Q}^{1/2}\Upsilon$.
To solve this combinatorial optimization,
we adopt
an iterative greedy algorithm
called {\em Orthogonal Matching Pursuit} (OMP)
\cite{PatRezKri93}.
The algorithm is very simple and significantly faster than
exhaustive search.
Moreover, OMP always gives a vector in the feasible set of the optimization \eqref{eq:opt0},
and hence the networked control system will be asymptotically stable by Theorem~\ref{thm:stability}.

\section{Example}
\begin{figure}[t]
\centering%
\mbox{
\subfigure[Performance $\|\vx(k)\|_2$]{\includegraphics[width=0.4\linewidth]{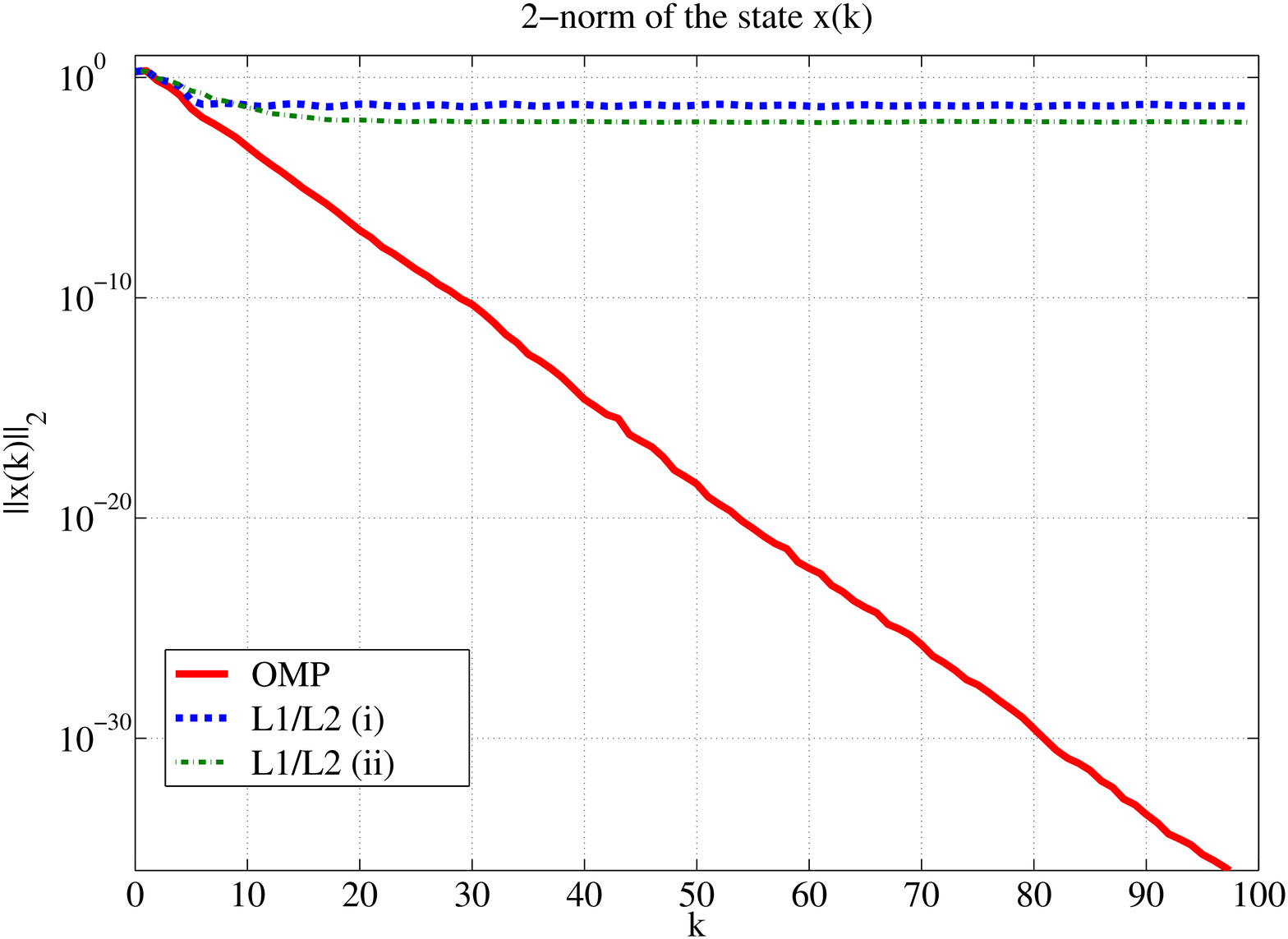}}
\subfigure[Sparsity $\|\vu(\vx(k))\|_0$]{\includegraphics[width=0.4\linewidth]{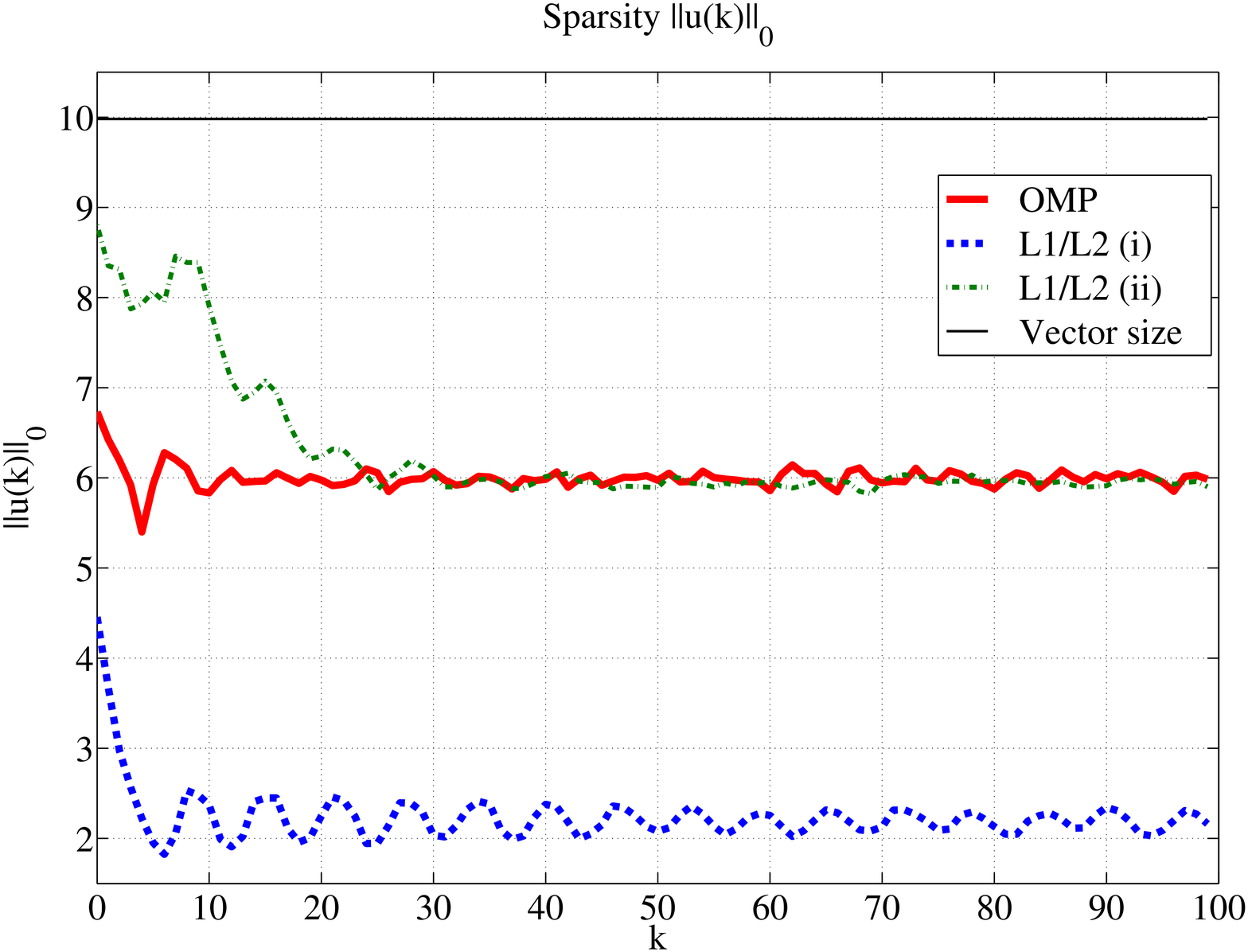}}
}
\vspace{-5mm}
\caption{Results by OMP (solid) and $\ell^1/\ell^2$ optimization (dash)}
\label{fig:results}
\end{figure}

To show the effectiveness of the proposed method, we run 500 simulations with
a fixed 4-th order unstable plant
(the poles are
  $-1.4396$,
  $1.0808 \pm 0.6664i$, and
  $0.0220$).
The packet size $N$ is 10.
We generate a packet dropout at each time $k$
with dropout probability $1/2$ (if there have been $N-1$ consecutive dropouts,
we set the next dropout probability to be $0$, i.e., 
the dropout process is Markovian).
Figure \ref{fig:results} compares  averaged results of the proposed method with that of \cite{NagQue11}.
Two plots (i) and (ii) are displayed for the method of \cite{NagQue11},
with different design parameters for sparsity.
We can see that the OMP control exhibits asymptotic (even exponential)
stability, whereas the $\ell^1/\ell^2$ 
method of \cite{NagQue11} is only practically stable.
The $\ell^1/\ell^2$ method produces much sparser vectors as in (i)
than the OMP formulation, but the system does is not asymptotically stable
even if one relaxes the sparsity constraint in the $\ell^1/\ell^2$ optimization
as in (ii).
In fact, it is proved in \cite{NagQue11} that if the controlled plant is unstable,
asymptotic stability is impossible by
the $\ell^1/\ell^2$ method.
The reason is that the $\ell^1/\ell^2$ method amounts to considering a fixed upper bound
for the inequality constraint in \eqref{eq:opt0} or \eqref{eq:opt0_2}
instead of time-varying $\vx(k)^\top W\vx(k)$.
This leads to $\vz$-valued control vector when $\|\vx(k)\|_2$ is sufficiently small
(note that $\vu=\vz$ is the sparsest vector among all vectors).

\end{document}